\documentclass[sigconf]{acmart}
\acmYear{2025}\copyrightyear{2025}
\setcopyright{cc}
\setcctype[4.0]{by}
\acmConference[SoCC '25]{ACM Symposium on Cloud Computing}{November 19--21, 2025}{Online, USA}
\acmBooktitle{ACM Symposium on Cloud Computing (SoCC '25), November 19--21, 2025, Online, USA}
\acmDOI{10.1145/3772052.3772256}
\acmISBN{979-8-4007-2276-9/25/11}
\usepackage{hyperref}
\hypersetup{breaklinks=true}

\usepackage{xspace}
\usepackage{amsmath}
\usepackage{graphicx}
\usepackage{siunitx}
\usepackage{caption}
\usepackage{fontawesome}
\usepackage{enumitem}
\usepackage{booktabs}
\usepackage[capitalize]{cleveref}
\usepackage[scaled=0.82]{FiraMono}

\usepackage[many]{tcolorbox}
\tcbset{
    sharp corners,
    colback = white,
	colframe=black!50!white,
    before skip = 0.2cm,    
    after skip = 0.5cm      
}                           
\usepackage{listings}
\newtcolorbox{boxK}{
    sharpish corners, 
    boxrule = 0pt,
    toprule = 4pt, 
    enhanced
}
\newcommand{\fffs}{\textsc{f\oldstylenums{2}fs}\xspace}

\newcommand{\btrfs}{\textsc{btrfs}\xspace}

\newcommand{\zonefs}{\textsc{z}one\textsc{fs}\xspace}

\newcommand{\zenfs}{\textsc{z}en\textsc{fs}\xspace}

\newcommand{\ssd}{\textsc{ssd}\xspace}
\newcommand{\Ssd}{\textsc{Ssd}\xspace}
\newcommand{\ssds}{\textsc{ssd}s\xspace}

\newcommand{\zns}{\textsc{zns}\xspace}

\newcommand{\fdp}{\textsc{fdp}\xspace}

\newcommand{\lsm}{\textsc{lsm}\xspace}

\newcommand{\uuid}{\textsc{uuid}\xspace}

\newcommand{\cpu}{\textsc{cpu}\xspace}

\newcommand{\posix}{\textsc{posix}\xspace}

\newcommand{\nvme}{NVMe\xspace}
\newcommand{\io}{I/O\xspace}
\newcommand{\api}{API\xspace}

\newcommand{\Shimmer}{Valet\xspace}
\newcommand{\vfs}{\emph{valet-mapper}\xspace}

\newcommand{\sysint}{syscall\_intercept\xspace}
\newcommand{\ldpreload}{\texttt{LD\_PRELOAD}\xspace}

\newcommand{\pathmap}{\texttt{pathMap}\xspace}
\newcommand{\fdmap}{\texttt{fdMap}\xspace}
\newcommand{\fdstream}{\texttt{streamMap}\xspace}
\newcommand{\filemap}{\texttt{fileMap}\xspace}
\newcommand{\zonemap}{\texttt{zoneMap}\xspace}

\newcommand{\wesdig}{Western Digital Ultrastar DC ZN540\xspace}

\makeatletter
\def\verbatim@font{\small\ttfamily}
\makeatother

\begin{document}
\frenchspacing
\title{\Shimmer: Efficient Data Placement on Modern SSDs}

\author{Devashish R. Purandare}
\email{dpuranda@ucsc.edu}
\orcid{0000-0001-9434-3933}
\affiliation{%
	\institution{UC Santa Cruz}
	\city{Santa Cruz, CA}
	\country{USA}
}
\author{Peter Alvaro}
\email{palvaro@ucsc.edu}
\orcid{0000-0001-6672-240X}
\affiliation{%
	\institution{UC Santa Cruz}
	\city{Santa Cruz, CA}
	\country{USA}
}
\author{Avani Wildani}
\email{agadani@gmail.com}
\orcid{0000-0001-9457-8863}
\affiliation{%
	\institution{Emory University / Cloudflare}
	\city{San Francisco, CA}
	\country{USA}
}
\author{Darrell D. E. Long}
\email{darrell@ucsc.edu}
\orcid{0000-0002-0822-0740}
\affiliation{%
	\institution{UC Santa Cruz}
	\city{Santa Cruz, CA}
	\country{USA}
}
\author{Ethan L. Miller}
\email{elm@ucsc.edu}
\orcid{0000-0003-2994-9060}
\affiliation{
	\institution{Pure Storage / UC Santa Cruz}
	\city{Santa Cruz, CA}
	\country{USA}
}

\begin{abstract}
	The increasing demand for \ssds coupled with scaling difficulties has left
	manufacturers scrambling for newer \ssd interfaces which promise better
	performance and durability. While these interfaces reduce the rigidity of
	traditional abstractions, they require application or system-level changes
	that can impact the stability, security, and portability of systems. To make
	matters worse, such changes are rendered futile with the introduction of
	next-generation interfaces.  It is therefore no surprise that such
	interfaces have seen limited adoption, leaving behind a graveyard of
	experimental interfaces ranging from open-channel \ssds to stream \ssds.

	Our solution, \Shimmer, leverages userspace \emph{shim layers} to add
	placement hints for application data, delivering up to 2–4$\times$ write
	throughput over filesystems and comparable or better performance than
	application-specific solutions, with up to 6× lower tail latency.  \Shimmer
	generates dynamic placement hints, remapping application data to modern
	\ssds with \emph{zero modifications} to the application, the filesystem, or
	the kernel.  We demonstrate performance, efficiency, and multi-tenancy
	benefits of \Shimmer across a set of widely-used applications: RocksDB,
	MongoDB, and CacheLib, presenting a solution that combines the performance
	of application-specific solutions with wide applicability to log-structured
	data-intensive applications.

\end{abstract}



\maketitle

\section{Introduction}


As the demand for datacenter \ssds soars, scaling flash has become increasingly challenging,
with density increases adversely impacting performance and device
lifetime~\cite{grupp2012bleak}. To make matters worse, \textsc{nand}-flash \ssds cannot
perform in-place updates and have large erase units, requiring valid data
relocation to free up space. These garbage collection operations impact
performance and device lifetime due to internal data movement and write
amplification. Even log-structured, append-only systems, which are designed to
match device characteristics, are limited by traditional interfaces and fail to
reach their full potential. With application logs, misaligned on filesystem
logs, further misaligned on device logs, the “log-on-log
problem”~\cite{yang-inflow14} causes performance and lifetime degradation with
redundant garbage collection at multiple levels.

In this paper, we focus on log-structured storage patterns as they make up almost
all modern data-intensive systems, owing mainly to the performance benefits on
hard disks (reduced seek), and efficiency benefits on \ssds (reduced garbage
collection). Append-only log-structured storage backends are used by all major
cloud storage providers~\cite{calder2011windows, verbitski2017amazon,
    hildebrand2021colossus}, universally used in data-intensive
applications~\cite{bookkeeper2024, pulsar2024, spanner2024,cassandra2024}, and
ubiquitous in modern filesystems~\cite{lee2015F2FS,bonwick2003zfs,
    rodeh2013btrfs}.

Two key placement strategies can help log-structured systems take full advantage
of \textsc{nand} flash. First, a \emph{lifetime-aware placement} that clusters
data with a similar lifetime (i.e., whose creation and deletion have temporal
locality) can minimize data relocation on erase, improving overall performance
as well as device endurance. Second, an \emph{affinity-aware placement} that
clusters data produced by a single application together and places independent
data streams on separate device resources can provide performance isolation,
reducing latency spikes, which have a major impact on cloud storage
performance~\cite{10.1145/2408776.2408794}.

These strategies can (in principle) be realized in the current state of the art:
flash manufacturers have introduced storage interfaces such as Zoned Namespace
SSDs (\zns)~\cite{230004,purandare-cidr22}, and Flexible Data Placement
(\fdp)~\cite{sabol_stenfort}, which allow the host to direct data placement on
the \ssd via \emph{placement directives} of a variety of forms, ranging from
placement hints at one extreme to assuming the responsibility for placement and
garbage collection at the other. However, such changes require using newer
abstractions or retrofitting current interfaces.

But exactly which layer should handle such abstraction-breaking
host-device coordination features? One answer is the application itself, which
knows best about its own data access patterns. This approach is fragile and
risky. Implementing device-specific optimization in application space requires
specialized expertise, and these efforts are likely to be made obsolete in the
face of \api changes. The other answer is the filesystem, which suffers from the
opposite problem. While implementing support for device-specific placement
directives in the operating system would shield applications from complexity and
provide much better generality and reuse, filesystems provide
interfaces that are too narrow to take advantage of application-level semantics
with regard to lifetime and affinity. The cost of generality means failing to
take advantage of device characteristics.

In this work, we argue that the responsibility for exploiting emerging
device-side placement directives should fall neither on the application
programmer nor on the kernel developer, but rather on the users or systems that
understand the \emph{application-device mapping}. Typically, neither application
developers nor filesystem developers are aware of the exact storage architecture
their work will be deployed on, and hence cannot make effective optimization
decisions. For every application-device combination, there is a space of possible
placement directives that may be factored apart from the application and
filesystem. Strategies to optimize placement based on lifetime and affinity
(application properties) utilizing the features provided by devices (from hints
to full management of regions) can be expressed in a separate \emph{shim layer}
that interposes between application and OS. Many configurations (indeed, the
product of devices and applications) must be implemented, but these are easier
to implement in isolation than within applications or operating systems.  By
isolating the complexity of each interface in a module decoupled from applications
and filesystems, we can shield all other layers of the system from change.

Our shim layer approach opens up a generalizable interface that is
application-agnostic, but can be optimized per application. It isolates the
complexity of varying interfaces and hint generation in pluggable userspace
modules, allowing quick and easy changes to work with changing interfaces. Our
library, \Shimmer, presents a blueprint for dealing with the complexity of
host-device coordination, allowing dynamic placement decisions without requiring
application or filesystem rewrites. Our key insight is that shim layers
can offer the \emph{performance} of custom solutions, and \emph{compatibility}
of filesystems, while reducing the \emph{complexity} of using modern interfaces.

\subsection*{Our contributions:}
We demonstrate how shim layers can provide both performance and compatibility
while reducing application and filesystem complexity.  We evaluate \Shimmer on
three widely used applications (RocksDB, MongoDB, and CacheLib) across two types
of interfaces (\zns and kernel hints); more applications and interfaces than any
previous effort.  To our knowledge, this is the first work to present a
generalized theory of data placement, showcasing \emph{affinity} and
\emph{lifetime} as the important parameters over temperature-based approaches of
the past.  We deploy \Shimmer with heuristic and learning-based hints,
showcasing extensibility which is difficult to achieve in filesystems or
applications.  \Shimmer is fast: we see 2–6 times higher write throughput, up to
6 times lower latency, and reduced garbage overhead over filesystems and
application backends.

\section{Benefits of host-guided data placement}\label{section:whyshim}

Hosts can greatly alleviate the need for garbage collection on \ssds and
provide performance isolation for data streams if they can communicate data grouping
to the device. This prevents interleaving of unrelated data on flash and reduces
drive fragmentation. However, as we see in~\cref{table:bleak}, manufacturer
demonstrations of these interfaces have focused on 1–2 applications due to the
complexity of adapting data-intensive applications to abstraction-breaking
changes.  To motivate the need for techniques that offer wider compatibility
than application-specific approaches while maintaining their performance, we
show a simple experiment:

\begin{table*}[htbp]
	\caption{Data placement abstractions over the years provide a sobering
		reality: with new interfaces demonstrating a few applications before being
		deprecated for lack of use.}
	\label{table:bleak}
	\centering
	\begin{tabular}{l l l l l }\toprule
		\textbf{Type of \Ssd}          & \textbf{Multi-Stream} & \textbf{Open-Channel} & \textbf{Zoned Namespaces} & \textbf{Flexible Data Placement} \\ \midrule
		\textbf{Introduced}           & 2016                  & 2017                  & 2021                      & 2022                             \\
		\textbf{Linux Kernel Support} & Deprecated            & Deprecated            & Yes                       & —                                \\
		\textbf{Hint Interface}       & \texttt{fcntl()}      & \texttt{liblightnvm}  & \texttt{libzbd,libnvme}   & \texttt{libnvme}                 \\
		\textbf{Filesystems}          & —                     & —                     & \fffs, \btrfs             & —                                \\
		\textbf{Applications}         & AutoStream            & RocksDB               & RocksDB                   & Cachelib, RocksDB                \\ \bottomrule
	\end{tabular}
\end{table*}


Using a \SI{4}{\tera\byte} \wesdig \ssd~\cite{wd_ultrastar_dc_zn540}, we
performed sequential write tests with flexible \io tester
(\texttt{fio})~\cite{fio2020} scaling up to 14 threads (the maximum open zones
supported by the drive). We ran the tests on \zonefs~\cite{246544}, a
block-layer representation of the \zns interface, and \fffs~\cite{lee2015F2FS},
a flash-optimized filesystem. We made sure that the writes on \zonefs went to
different zones, while on \fffs we provided the \emph{sequential} write and the
\emph{extreme} lifetime hints. To ensure parity, we used Direct \io, instructing
\fffs to skip the buffer cache (\zonefs does not support write buffering).

\begin{figure*}[htbp]
	\centering
	\includegraphics[width=\textwidth]{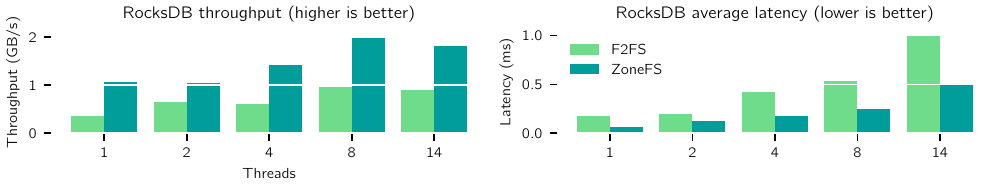}
	\caption{Writes to \zonefs can get the full bandwidth while \fffs sees
		degradation in both latency and throughput.}
	\Description{A dual-panel graph showing throughput and latency for zonefs
	and F2FS across 1-14 concurrent writers. Zonefs maintains linear throughput
	scaling and low latency, while F2FS throughput plateaus around 4 writers
	with increasing latency.}
	\label{fig:hintbenefit}
\end{figure*}

The results~\cref{fig:hintbenefit}, show that a lightweight mapping layer with
the right hints (map each file to a separate zone) can provide full device
throughput, while \fffs is limited to 30–50\% of the bandwidth. We analyzed the
results in \texttt{perf}~\cite{de2010new} and break them down by the
\textsc{cpu} cycles spent by our test in each scenario. Even with
\texttt{O\_DIRECT}, \fffs needs to cache writes to map them to various segments.
Such caches result in frequent internal data structure updates and syncs. This
overhead adds up with in-kernel locking operations resulting in \fffs spending
more time in sync (34.43\%) than in write calls (12.72\%). The added overhead
results in 2–3$\times$ higher latency and lower throughput. While in \zonefs,
since the filesystem is aware that these are writes to separate zones, it does
not need to cache or sync to the device, utilizing the full bandwidth of the
device buffer.

However, \fffs is a full-fledged filesystem while \zonefs is closer to a raw
block device. Such overhead imposed by filesystems can be greatly reduced by
designs which are aware of the log-structured nature of incoming data as well as
the underlying device.

\begin{figure}[htbp]
	\centering
	\includegraphics[width=\columnwidth]{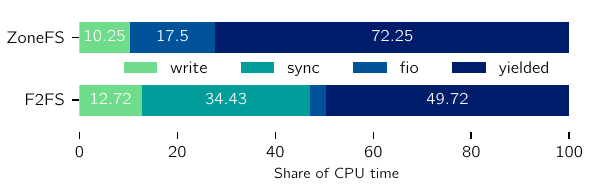}
	\caption{CPU time breakdown shows \zonefs yields 72\% of CPU time back while \fffs only yields 49\%, with 34\% spent on synchronization overhead.}
	\Description{Bar chart comparing CPU time usage between F2FS and ZoneFS. F2FS yields 49\% of CPU time with 34\% consumed by synchronization overhead. ZoneFS yields 72\% of CPU time by avoiding synchronization overhead entirely.}
	\label{fig:breakdown}
\end{figure}

\subsection*{Picking the right layer for coordination}

Traditionally, developers have implemented support for host involvement either
at the application or the filesystem. We argue that both these approaches have
major limitations.

\subsubsection*{Why not rewrite the application?}
\begin{enumerate}[leftmargin=*]

	\item \emph{Rewriting applications is expensive:} Rewriting applications for
	a specific architecture requires significant engineering effort. For
	instance, the \zenfs~\cite{bjorling2021zns} project, which optimizes RocksDB
	for \zns, is a multi-year effort with thousands of lines of code, which has
	been largely abandoned, with only a single commit since October
	2023~\cite{zenfs_github_issue_288}.

	\item \emph{Applications are storage-unaware:} Applications typically use the
	      file interface and are abstracted from the nature of storage. They are unaware
	      of system usage or other applications, resulting in any efforts at resource
	      acquisition and hinting being ineffective.

	\item \emph{Modifying applications limits layering and portability:} Even
	      if an application is customized end-to-end to use custom interfaces, it
	      cannot then effectively address multiple types of devices. Breaking away
	      from the file abstraction can hurt tooling for operating on the data like
	      replication and backup utilities.

\end{enumerate}

\subsubsection*{What about a filesystem?}\label{sec:why-not-fs}
\begin{enumerate}[leftmargin=*]
	\item
	      \emph{Filesystems are application-unaware:} Mainstream filesystems are
	      designed to be independent of the applications running on top of them
	      and find it difficult to generate useful hints unless communicated by
	      the application. Currently, no standard interfaces exist: for example,
	      \fffs can utilize hints from the multi-stream \ssd interface and map it
	      to zones, but just three hint levels are insufficient across all applications.

	\item
	      \emph{Filesystems are hard to modify:} As filesystems reside in the kernel,
	      they are hard to modify and upgrade. Fixing \zns-related bugs in \fffs, for
	      instance, requires upgrading to a newer version of the Linux kernel, which
	      is impractical for data centers as it requires migration and downtime and
	      may cause issues with other applications. Adding complexity to the kernel
	      can give rise to crashes and security vulnerabilities.

	\item \emph{Filesystems require broad compatibility:} Adapting a
	      filesystem for \zns, for instance, should not limit support for other
	      types of \ssds. The increasing complexity can result in increased bugs
	      in the kernel, reduced performance, and an increased attack surface.

\end{enumerate}
While some of these issues could be addressed, for example with a
FUSE~\cite{vangoor2017fuse} filesystem, it would still require a
per-architecture per-application filesystem to utilize host-device hint
mechanisms fully. Such usage of FUSE would be similar to our proposed shim layer
but with the added complexity of managing per-application filesystems. Further,
repeated kernel crossings communicating between modules, mappings, and hint
generation could negate any performance benefits from modern storage interfaces.

Not only do extra interfaces to filesystems increase bugs, due to their
in-kernel nature they add security vulnerabilities. Rather than adding bloat to
the kernel, we can isolate the complexity in a small audit-able layer, greatly
reducing attack surface while improving performance.

\subsubsection*{The Middle Road: Shim Layers}
A shim layer can abstract interface changes from the applications and
filesystems while enabling low-overhead reconfiguration to exploit the benefits
of modern \ssds. In this architecture, simplicity is maintained in the
application and the filesystem, while the added complexity of hinting is
isolated in a configurable layer—enabling low-cost, relatively-low-effort
adoption.

With the goal to allow efficient adoption of modern \ssds, an ideal shim layer should require:
\begin{enumerate}[leftmargin=*]
	\item \emph{No changes to the applications or operating system:} To simplify
	      adoption, a shim layer should not need any changes to the application, any
	      kernel modules, or reconfiguration of the system.
	\item \emph{Broad compatibility:} The shim layer should be able to work
	      across different applications and utilities.
	\item \emph{Efficiency and effectiveness:} A shim layer should unlock
	      performance benefits without adding more overhead than a tuned application
	      or filesystem.
	\item \emph{Extensibility:} The hint generation should be configurable, adding
	      the ability to add custom logic, including systems that learn dynamically.
\end{enumerate}

\noindent
We built \Shimmer to stay true to these principles, and we demonstrate that such
a layer is not only feasible; it can outperform other approaches. With \Shimmer,
we propose the \emph{addition of a layer} to break traditional layering
abstractions, as it can \emph{isolate changes across layers} without impacting
the compatibility and portability of the application.

\section{\Shimmer Architecture}
\begin{figure}[htbp]
	\centering
	\includegraphics[width=\columnwidth]{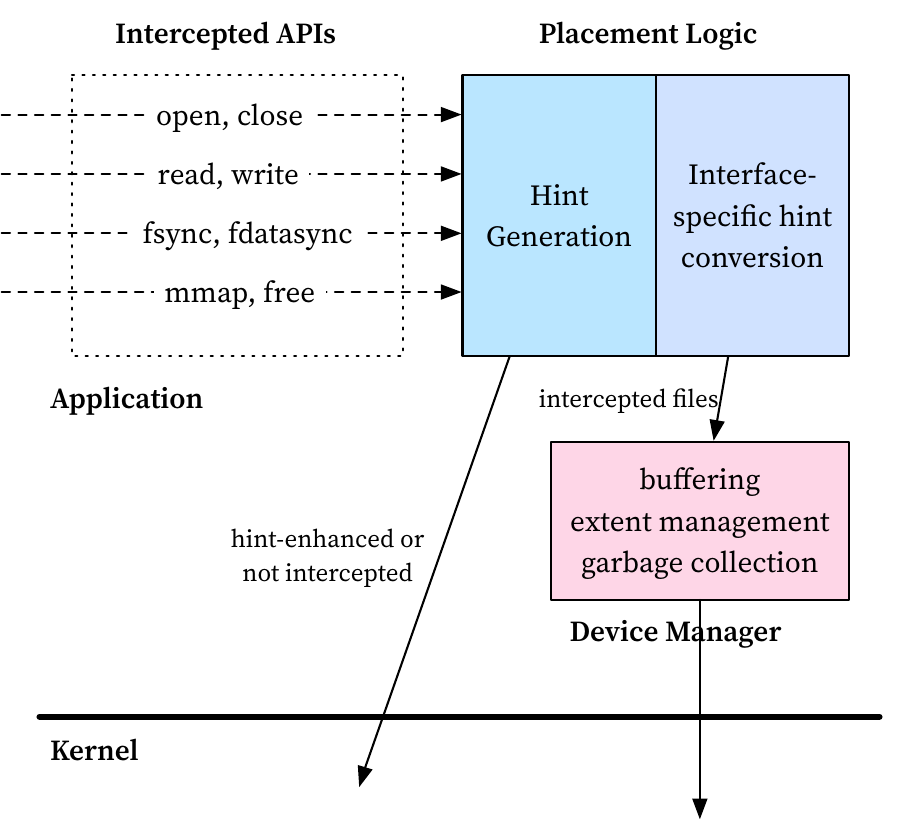}
	\caption{Simplified \Shimmer architecture: \Shimmer intercepts application
		calls, generates placement plans, resolves them to a particular protocol and
		finally manages data placement.}
	\Description{Architecture diagram showing system call flow from application through Shimmer to storage device. Shimmer intercepts system calls at the Call Interception layer, which redirects them through the Data Placement Engine for hint generation, then to the Device Manager for physical placement. Some calls bypass interception and pass directly to the underlying filesystem. The three-component pipeline connects applications to ZNS SSDs or hint-based storage interfaces.}
	\label{fig:shimmer}
\end{figure}

\Shimmer is a dynamic library that allows interposition on application calls,
modifying them if necessary, to embed placement directives or redirect them to
different regions on an \ssd.

\Shimmer allows transparent hint injection or redirection of data based on the
decision made by the placement engine. Since \Shimmer has insight into the
application and the storage architecture, it can generate more effective hints
than applications or filesystems in isolation. Further, users can modify the
hint generation easily without having to rewrite, reconfigure, or recompile the
application or the filesystem.

\Shimmer performs three tasks necessary for host-device coordination, which act
at different layers of the modern stack—the application, the storage engine,
and the device. Due to the differing hardware design approaches, we designed each
of these components to be customizable and pluggable with configuration changes.
For instance, in the \zns protocol, the device is partitioned into equal-sized
append-only zones. The host is responsible for picking zones, managing buffers
and garbage collection when it needs to free up space. This approach requires
the \emph{device manager} which is implemented in \vfs.

Approaches like multi-stream and Flexible Data Placement (\fdp) on the other
hand, simply require a directive on where to place the data, hence the device
manager can be skipped for those devices.

For \Shimmer we present implementations for both of these approaches with
demonstration using \zns drives with \wesdig and the ability to use kernel hint
interfaces for multi-stream \ssds.  Currently, \fdp support is limited: hardware
is not generally available and the directives (by design) are only supported
through \nvme \io.  Since most applications do not use raw \nvme \io (outside of
exceptions like CacheLib~\cite{cachelib}), we plan to look at \fdp support as
future work. This could be achieved by extending \vfs to use kernel \nvme device
support and adding an extent mapping layer.

\noindent As seen
in~\cref{fig:shimmer}, \Shimmer has 3 major components:
\begin{enumerate}[leftmargin=*]
	\item {Application call interception:} \Shimmer intercepts system
	      calls issued by the application.
	\item {Data Placement Engine:} \Shimmer uses automated or heuristic
	      placement logic to group data and generate placement hints.
	\item {Device Manager:} Depending on the device type, \Shimmer can
	      either forward the hints or perform the physical placement.
\end{enumerate}

\subsection{Application Call Interception}
\label{sec:interception}
\begin{table*}[htbp]
	\centering
	\caption{\ldpreload provides a valuable compromise, requiring no changes to
	the application or the kernel.}
	\label{table:compare}
	\begin{tabular}{lllllll}\toprule
		                            & \textbf{Added Kernel} & \textbf{Static}  & \textbf{Kernel}  & \textbf{Per-Application} & \textbf{Shim}  & \textbf{Performance}                         \\
		                            & \textbf{Crossings}    & \textbf{Linking} & \textbf{Changes} & \textbf{Configuration}   & \textbf{Setup} & \textbf{Degradation}                         \\ \midrule
		\texttt{LD\_PRELOAD}        & –                     & –                & –                & Yes                      & Runtime        & Minimal\footnotemark                         \\
		\textbf{eBPF}               & Yes                   & Yes              & Yes              & Yes                      & In Advance     & $\sim$10$\times$~\cite{zheng_bpftime_2023}   \\
		\textbf{WASM}               & –                     & Yes              & –                & Yes                      & In Advance     & $\sim$2.5$\times$~\cite{jangda_not_nodate}   \\
		\textbf{FUSE}               & Yes                   & Yes              & Yes              & –                        & In Advance     & $\sim$1.8$\times$~\cite{vangoor2017fuse}     \\ \bottomrule
	\end{tabular}
\end{table*}

Shim techniques have seen a renaissance as fast-changing hardware,
special-purpose processing, and changed memory hierarchies become common in our
systems. It is no surprise therefore that we see a proliferation of techniques
that allow shimming \texttt{libc} calls like eBPF~\cite{bijlani2019extension},
FUSE~\cite{mizusawa2017performance}, WASM~\cite{wasi-overview}, virtual
machines, and dynamic libraries~\cite{noauthor_syscall_intercept_2023} being
widely deployed. While \Shimmer can be implemented with any of these techniques,
we focused on \ldpreload to avoid kernel crossings, allow us to
implement an all-userspace system and minimize performance degradation.
Dynamically linking with the application allows us to modify \texttt{libc} calls
before they are sent through the kernel, allowing a simple \io path through the
kernel, while the complexity resides in userspace. \ldpreload is used by
projects ranging from custom allocators to debugging tools~\cite{jemalloc,
	valgrind}.

\cref{table:compare} presents the design space of shim techniques. We chose the
one with the best performance, and while its limitations may exclude certain
applications, the principles we discuss could be realized with other techniques
or even a custom \texttt{libc}.  Simplicity of implementation, userspace nature,
and runtime linking make our approach easy to use, requiring \emph{no
recompilation, no changes to the application, and no changes to the system}. We
discuss the limitations of our approach and alternatives in
\cref{sec:limitations}.

\Shimmer is designed to be dynamically linked with the application at runtime
using \ldpreload, which allows the library, \texttt{libvalet}, to be loaded
before other objects while executing a binary. \Shimmer uses this functionality
to selectively override various \texttt{libc} calls, using \Shimmer calls
instead.  In hint-based interfaces, \Shimmer injects placement directives
received from the placement engine.  Here hint-based approaches like \fdp or
multi-stream \ssds require just a placement identifier, while host-managed
approaches like \zns require the host to take over data placement, garbage
collection, and resource management. For hint-based approaches, \Shimmer works
on top of an existing storage system that supports hints like \fffs.  For
host-managed approaches, \Shimmer implements its own device manager with \vfs
and provides placement information per file. When used with \vfs, \Shimmer
captures the content of calls the application issues, copying the data to its
buffer cache before persisting it based on hints.

In both cases, \Shimmer needs to maintain a certain level of bookkeeping to
supply directives per file.  With the system calls that use file descriptors which
are ephemeral and issued per-session, \Shimmer needs to maintain a mapping of
files to their descriptors to effectively understand which file a call is
operating on. Since \Shimmer does not have a runtime independent of the
application, it leverages library load to set up a multi-threaded state, and
specific conditions to request the read, write, garbage-collection or metadata
threads to perform a task.


When working on top of a filesystem, \Shimmer issues the \texttt{open()} system
call to the underlying filesystem, getting the kernel-issued file descriptor
(\texttt{fd}), and then assigns a unique identifier (\uuid) to the file,
maintaining two mappings: \texttt{path→uuid} (the \pathmap) and \texttt{uuid→fd}
(the \fdmap) in separate hash tables. The \fdmap is automatically updated on
each \texttt{open()} and \texttt{close()} call and never persisted, while the
\pathmap is updated on creates (seen through \texttt{open()}, \texttt{rename()},
and \texttt{unlink()}) and persisted on sync operations.  Data-intensive
applications are often multithreaded and can result in separate threads
accessing the same file descriptor concurrently. This can quickly lead to
contention on lookups, especially as \Shimmer is invoked on an
intercepted call. To avoid frequent locking, \Shimmer uses
\texttt{dashmap}~\cite{dashmap}, a concurrent hashmap, for its state.

\footnotetext{\ldpreload and \texttt{syscall\_intercept} added
	$\sim$\SI{2}{\micro\second} and $\sim$\SI{2}{\milli\second} of userspace
	overhead per syscall in our tests respectively.}

\Shimmer's interception varies depending on whether the interface is hint-based
or host-managed. Hint-based approaches are much simpler, but they lose out on
some efficiency gains:

\subsubsection{\Shimmer in hint-based systems}
In hint-based interfaces like multi-stream and \fdp, \Shimmer does not need \vfs, so its
interception requires limited state tracking, operating once per created file.

\begin{enumerate}[leftmargin=*]
	\item \texttt{library-load:} On library load, \Shimmer blocks and sets up
	necessary state including loading previous metadata, updating internal data
	structures, read, write, garbage collect, and metadata threads, and loads
	the hint-generation configuration or model. This allows us to avoid
	allocation and initialization in hot data path.

	\item \texttt{open()}: If a new file is opened in write mode, \Shimmer
	provides the path and open flags to the hint generation logic, adds the file
	to tracked files (\fdmap and \pathmap), and issues the necessary
	\texttt{fcntl()} and \texttt{fadvise()} calls with the hints returned by the
	placement engine.

	\item \texttt{close()}: removes the entry associated with the \texttt{fd} in
	\fdmap and syncs persisted \pathmap before forwarding \texttt{close()} to
	the filesystem.

	\item \texttt{unlink()}: \Shimmer updates both maps to remove the \uuid on a
	successful return of the unlink call from the underlying filesystem.

	\item \texttt{fsync()}: With \texttt{fsync} and its variants
	(\texttt{fdatasync},\\ \texttt{sync\_file\_range}), \Shimmer syncs \pathmap
	on a successful return from the underlying filesystem.

	\item \texttt{rename()}: On a successful rename, \Shimmer updates \pathmap
	with the new path.
\end{enumerate}

\subsubsection{\Shimmer in host-managed systems} Host-managed interfaces like
\zns require deeper support than issuing just the required placement hint, and
we implement \vfs, a lightweight device management engine to enable management
of flash.

Some setup tasks are similar between hint-based and host-managed interfaces,
but in the case of host-managed interfaces, \Shimmer needs to ensure more than
simply issuing the right hint. It needs to notify \vfs to perform the needed
task, not dissimilar to a lightweight virtual filesystem.

\begin{enumerate}[leftmargin=*]
	\item \texttt{library-load:} The startup is similar to hint-based logic, but
	in addition to restoring states, \Shimmer spins up \vfs, allowing it to
	allocate its metadata structures and pre-allocate write buffers for the
	buffering logic.

	\item \texttt{open()}: Open performs the same tasks as described previously,
	however, in the host-managed case, instead of issuing a hint system call,
	\Shimmer requests a stream from the hint generation logic, mapping the
	current files to the stream as discussed in
	\cref{sec:hint-generation,sec:dm}.

	\item \texttt{close()}: removes the fd from \fdmap and \fdstream. Requests
	\vfs to flush data associated with the \texttt{fd}. \vfs further syncs its
	own metadata on every close (to uphold the \posix contract on data persistence).

	\item \texttt{unlink()}: Removes all references to the file, requests \vfs
	to mark associated data for cleanup. Sends a message to the garbage
	collection thread to check if it can free up space.

	\item \texttt{write()}: On the write call and its variants, \Shimmer
	forwards the buffer to \vfs with the \uuid which handles writing the data to
	the device, on a successful return from \vfs, it returns success to the
	application.

	\item \texttt{read()}: On the read call and its variants, \Shimmer
	translates the read to \uuid and offset before forwarding the request to
	\vfs, and returns the data it gets back.

	\item \texttt{fsync()}: On sync, \Shimmer persists its own mapping and
	forwards the request to \vfs to ensure the buffered data associated with the
	file is persisted.

	\item \texttt{rename()}: is unchanged from hint-based logic.
\end{enumerate}

In addition to the previously discussed calls \Shimmer needs to modify other
calls to guarantee persistence, prevent extra allocation, and maintain a
consistent state. \Shimmer performs the following actions on each of these
calls:

\begin{itemize}[leftmargin=*]
	\item \texttt{fallocate()}: In host-managed mode, these calls are suppressed as
	      \vfs uses a \emph{flush on sync} optimization.
	\item \texttt{ftruncate()}: In host-managed mode, these calls only update
	the metadata as in-place updates on host-managed interfaces are not allowed
	unlike hint-based \ssds.
	\item \texttt{readahead()}: Asks \vfs to readahead for the given range into
	its buffers.
	\item \texttt{mmap()}: this call is unsupported for host-managed mode
	(outside read-only open), as host-managed devices cannot support the
	in-place updates made by the call. \Shimmer forwards these calls to an
	in-place update friendly filesystem in the random-write area of the device.
\end{itemize}

As we see with \texttt{mmap()}, \Shimmer allows specific files not to be
intercepted using the hint mechanism. \vfs uses this for files that require
in-place updates—manifest and configuration files, as well as special purpose
files like \texttt{LOCK} files, device files, and \texttt{procfs} entries. This
ensures that applications work with minimal modifications and do not get any
unexpected errors, utilizing the default path for anything not implemented by
\vfs (e.g., \texttt{dup()}, \texttt{rmdir()}, etc.). Since most log-structured
systems need a small amount of in-place updatable files, \vfs maps them to
in-place update-friendly conventional zones and puts the log-structured data,
which makes up most of the data by volume, on sequential zones. Typically, these
files make up less than 1\% of the operations in log-structured systems.

Once the calls are intercepted, \Shimmer requests hints from the placement
engine. This pluggable module can either generate hints based on a heuristic or
an automated model.

\subsection{Data Placement Engine}\label{sec:hint-generation}
For a \emph{good} placement plan, data relationships need to be considered based
on two important properties:
\begin{enumerate}[leftmargin=*]
	\item{Data Affinity:} Semantically-related writes grouped together
	      maximize bandwidth through isolation.
	\item{Lifetime grouping:} Data that shares a common lifetime should be grouped
	      together to minimize free space fragmentation.
\end{enumerate}

For best results, a good placement engine must separate independent write
streams from across applications and within applications (such as data logs,
write-ahead logs, checkpoints, and manifests) into separate groups. Such
grouping will eliminate the interleaving of streams on device buffers and flash,
improving performance due to device-level isolation and parallelism.
Additionally, the system must group data by its \emph{expected} lifetime,
reducing the garbage collection overhead.

A major issue with implementing support for host-guided placement is the lack of
usable Kernel abstractions. The \texttt{RWH\_HINT} interface, leftover from
multi-stream \ssd days, allows four separate data streams: hot, cold, warm, and
undefined. So far, only one application (RocksDB) uses them, and they are
supported on a single filesystem (\fffs). There are several issues with this
approach: (1) three hint levels are insufficient with multiple applications, and
different data streams, and (2) temperature of data may not always be correlated
with lifetime.
These rigid interfaces mean effective placement requires custom filesystems or
kernel bypass.  \Shimmer, on the other hand, implements a
flexible internal hint representation, with resolvers for translating these
hints into currently supported APIs, and ability to extend support to future
APIs.  \vfs demonstrates the utility of richer interfaces and can unlock the
full potential of the placement logic (see \cref{sec:evaluation}).



\subsubsection{Hints in \Shimmer}

\begin{figure}[htbp]
	\centering
	\includegraphics[width=\columnwidth]{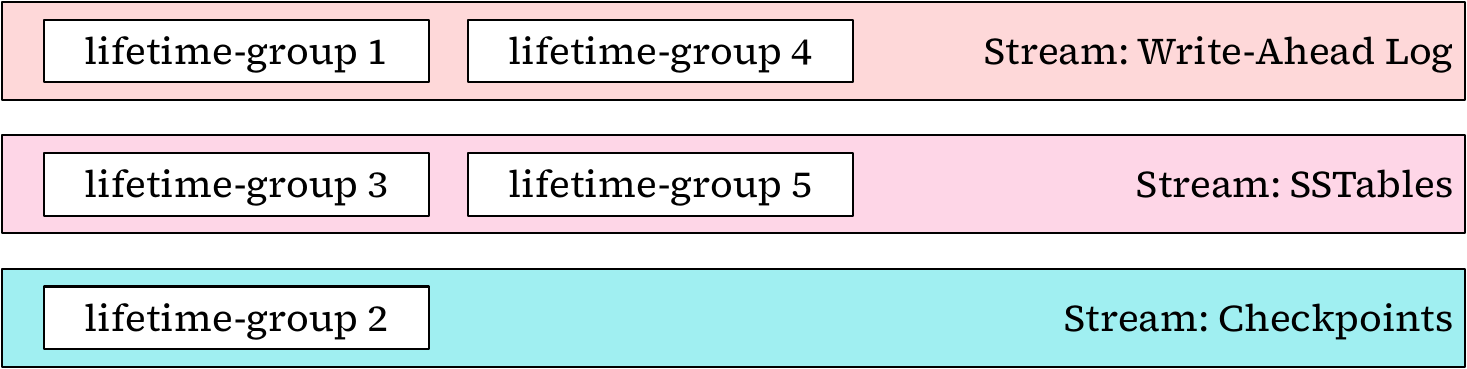}
	\caption{\Shimmer hints are organized based on affinity (streams) and
	lifetime (groups).}
	\Description{Diagram showing three parallel streams representing different data types: write-ahead logs, checkpoints, and SSTables. Each stream contains multiple numbered groups (lifetime groups) shown as boxes, illustrating the two-dimensional hint organization with streams for affinity-based separation and groups for lifetime-based clustering.}
	\label{fig:hint}
\end{figure}

Recognizing that grouping of data needs to be more than just a handful of
write-temperature streams, \Shimmer introduces two abstractions: \emph{streams}
and within them multiple \emph{lifetime-groups}. A stream represents a single
logical writer, which performs writes for a set of related data. This could be
writing to a write-ahead log, a data log, or performing compaction of existing
data. A lifetime-group on the other hand is a temporal subset of a stream,
grouping blocks within a stream which are written together. Streams are
application-specific, while lifetime groups are stream-specific.

As seen in~\cref{fig:hint}, an application could have different data streams: a
write-ahead log, data log (sorted string tables), and checkpoints.  Thus, unlike
traditional temperature-based hints, \Shimmer accounts for both affinity and
lifetime. We designed \Shimmer's internal hint representation to be simple and
extensible, allowing easy translation into the available hint formats and any
future changes.

\Shimmer's \texttt{get\_hint()} \textsc{api} is called with a resolver, which
can resolve hints into kernel hints, \vfs hints, or placement identifiers
depending on the configuration (see~\cref{sec:resolver}) allowing it to work
with existing hint systems as well as \vfs. Hint generation can be tuned per
application, these changes can be user-defined or automated. Streams can be
added by users familiar with application semantics and workload or generated
automatically based on observation.

\subsubsection{Providing Hints: Users vs. Automation}

For effective data placement, the placement engine needs to decide how to assign
streams for newly created files, and how to assign a lifetime-group for each
file.  As we focus on log-structured applications, we can assign the lifetime
group temporally, assigning a new group to blocks at fixed intervals of time or
space. As cleanup of the log happens, older data is compacted and written to the
tail of the log, allowing adjacent groups to be cleaned up together.

Distinguishing streams of incoming data from an application is more complex. We
demonstrate two techniques, tapping into heuristics and automation. For the
heuristic approach, we analyze the operation and the documentation of a
particular application and provide rules for sorting files. This approach is
particularly effective in applications like RocksDB which provide distinct
locations and file extensions for longer lived sorted string tables, and
short-lived write-ahead log.

If manual tuning is not preferred, hints can be
automated.  \Shimmer captures data such as paths, opening flags, and observed
workload, and can train on this information to dynamically predict streams. 
In \cref{sec:evaluation} we demonstrate this with batch-based mini
KMeans~\cite{10.1145/1772690.1772862} to dynamically pick streams based on
observed workload. \Shimmer-Learn picks the number of active streams that \vfs
prefers, and uses a kmeans clustering model to fit file open characteristics to
this set of streams. In the future, we plan to explore more dynamic features, as
well as splitting individual files across streams depending on the observed read
and write characteristics.

A placement algorithm that is independent of the application, filesystem, and
the architecture allows a lot of flexibility in tuning placement, where the
operators can use their expertise to tune hints based on their specific
architecture and workload.  In traditional systems such as applications or
filesystems, such flexibility is almost impossible, as the placement logic will
need to be in the kernel or within the application, and cannot be changed
without recompiling the entire system. If a filesystem allows dynamic placement
generation, it would need frequent kernel crossings to get placement
information, or worse, put the model in kernel-space.

\subsubsection{Resolving \Shimmer hints into interface hints}\label{sec:resolver}

To effectively use \Shimmer's hints across multiple interfaces, we implement
resolvers which can translate the internal representation into an
interface-specific hint. Translation to other formats can be lossy, especially
in cases like the Linux kernel hint interface (based on multi-stream \ssd),
which provides four values to choose from.
\begin{itemize}[leftmargin=*]
	\item \textbf{Multi-stream}: Here, we ignore the placement groups, assigning each
	      \Shimmer stream to kernel stream (hot, cold, warm, undefined), and in case
	      of more than four writers, we map multiple streams to each hint level,
	      maintaining a diminished level of write isolation.
	\item \textbf{Zones}: For \zns, we implement \vfs to demonstrate the full
	      potential of our approach. \Shimmer maps new zones to each stream
	      and ensures that placement-groups are laid out sequentially within a zone.
	\item \textbf{Placement identifiers}: Here, \Shimmer discard the streams,
	using \fdp placement identifiers with \Shimmer's lifetime-groups. As \fdp
	implementation is planned as future work, this resolver has not been
	implemented.
\end{itemize}

The advantage of decoupling hint interface from the application or filesystem is
that should a new hardware interface or software \textsc{api} be introduced, the
only change this logic would need is a new resolver to map internal
representation to the new format.  Armed with placement directives and
intercepted calls, we can now forward host-managed device support to \vfs,
another pluggable module to demonstrate \Shimmer's capabilities.

\subsection{Device Manager—the \vfs}\label{sec:dm}

As host-managed devices assign more responsibility (and hence, more power) to
the data management system on the host, we decided to implement a lightweight
storage engine to demonstrate the strengths of \Shimmer's approach.  With \zns,
\Shimmer's storage backend, \vfs takes over management of flash, presenting a
virtual filesystem interface to support reads and writes. Rather than writing
wrappers for device interfaces, we use \zonefs~\cite{246544}, a filesystem
wrapper in the kernel for zoned devices which allows using system calls rather
than \nvme commands to manage \zns \ssds. \vfs performs three main tasks: data
layout, buffering, and garbage collection.

\subsubsection{Data Layout}

Inspired by \zns' design, \vfs's data layout maps streams to individual zones.
To support arbitrary file sizes, \vfs implements an extent-based logic, allowing
block-aligned extents ranging from \SI{4}{\kibi\byte} to \SI{512}{\kibi\byte}.
On sync, the extent is padded to the nearest block-boundary and flushed to the
stream-mapped zone. Once the zone fills up, a new zone is fetched from a list of
free zones. \vfs uses \emph{allocate-on-flush}, deferring block reservation
until a close or sync command to further optimize grouping.


\vfs does not allow files to share chunks. This means that it only needs to
maintain the zone, the offset, and the size, to locate a particular extent. \vfs
maintains two data structures: the \pathmap to map paths to \uuid, and a
\filemap which maintains a file's extents as a list of \texttt{(zone, offset,
extent-size)} structures. \Shimmer maintains an additional structure, the
\zonemap, to maintain per-zone metadata to simplify garbage collection. \vfs
minimizes per-zone metadata by relying on the hardware write pointer to keep
track of offset, only maintaining a bitmap of the deletion status of all extents
on a zone. Effectively, the \zonemap requires 12–32 bits of memory per zone
depending on the extent size. Both \zonemap and \filemap are synced to
persistent storage on common operations like sync and close.

On \zns devices, since the allocation of device buffers is host-managed, \vfs
needs to track active resources and keep them below the device limits by
periodically finishing or closing zones. \vfs maintains a count of open zones
and closes them as they fill up. Since there can be a dozen or more open zones
at a time, this limit is rarely reached and typically happens when running
multiple applications in parallel.

\subsubsection{Buffering}

As \zonefs does not support write buffering, we implement write buffering in
\vfs. Due to the log-structured nature of applications handled by \vfs, we use a
simplified write-through buffer design. \vfs allocates a number of extent-sized
buffers on boot, maintaining them in a free list and assigning a new buffer to
each writable file. Once the buffer fills up, \vfs allocates blocks based on the
corresponding stream and flushes the buffer to storage. While this approach adds
a startup cost, we avoid allocating memory in the hot data path, improving
overall performance.


\subsubsection{Garbage Collection}


\vfs implements lazy garbage collection, deferring data movement as long as
possible. On each delete, \vfs updates the extent bitmap in the corresponding
zone in the \zonemap, and if all extents are marked as deleted, it resets the
zone, freeing up space without moving any data. In most log-structured systems,
frequently updated streams like write-ahead logs see frequent deletes, allowing
\vfs to reset fully-invalidated zones frequently. If available zones go below a
user-configured threshold, \Shimmer iterates through \zonemap identifying the
zone with the fewest valid extents and frees them up by moving the remaining
extents to new zones in the same stream.  Contiguous extents typically get
invalidated together in compaction operations.

\subsubsection{Crash Consistency}

\Shimmer offers similar guarantees as \posix filesystems on a crash.
Throughout execution, it performs careful metadata synchronization, updating
various persistent structures on calls to \texttt{fsync(), fdatasync()} or
\texttt{close()}, similar to filesystems. These updates write to a backup region
before an atomic commit, which swaps the current metadata with the backup.  This
ensures that metadata persistence failures do not corrupt the whole system.
Currently, we store the metadata in a human-readable JSON format on disk, as it
is relatively small (a few KiB) and helpful for offline data analysis. 

We made these decisions based on the general systems we want to replace with
\Shimmer, if there is a need for stronger guarantees or more efficiency, the
metadata module will require small updates to support write-ahead logging or
more efficient storage formats.  In the current design, a crash will cause data
between the crash and the last commit operation (triggered by sync or close
operations) to be lost. On a new boot, the constructor checks for metadata in a
known location on disk and reconstructs its state before allowing future
operations to proceed.

\subsection{Limitations}\label{sec:limitations}

\Shimmer inherits the limitations of \ldpreload that we discussed in
\cref{sec:interception}. While \Shimmer cannot intercept statically-linked
applications, in practice we observe that statically compiled applications like
RocksDB, Cachelib, and MongoDB still dynamically link with \texttt{libc}, and
can be preloaded with \Shimmer. The bigger limitation comes with languages that
do not use \texttt{libc} like Golang and Java, systems written in these
languages cannot be intercepted by \texttt{libc} replacements, and would need
one of the other shim approaches. Further, it is not always easy to preload the
library for complex client-server applications, as fork-execs and different
coordinating processes may spawn processes that lose the intercepted functions.
A different implementation of \Shimmer could use approaches that replace
\texttt{libc} like a custom library, or webassembly system interface to address
these applications.

We limit the scope of \vfs to log-structured append-only
applications. Since all major data-intensive systems almost exclusively follow
this pattern~\cite{10.1145/2043556.2043571,Bornholt2021,
53337,10.14778/3415478.3415546}, we can adapt several applications to this
interface, but applications with data structures that use in-place updates or
mmap writes cannot utilize \vfs and will be passed through to the filesystem on
conventional zones. \Shimmer can still issue hints for these writes if the
underlying filesystem supports them.

Finally, files stored by \vfs will not be visible to third party utilities like
backup and copy unless they are preloaded with \Shimmer as well. \vfs focuses on
data placement and is not a filesystem replacement as it does not implement
filesystem operations like access control or locking. We support these to a
limited extent on the random-write area on \zns drives by using a conventional
filesystem alongside \vfs. We could address these limitations by implementing
\Shimmer using other shim techniques, however, as we discussed
in~\cref{sec:interception}, each technique comes with its own set of trade-offs.
As we will see in~\cref{sec:evaluation}, we prioritized performance in our
design, and the system can be ported to other techniques for 
wider-applicability.

\section{Evaluation}\label{sec:evaluation}

To demonstrate the benefits of \Shimmer, we present three different types of
evaluation; we present three case studies with popular data management systems
RocksDB~\cite{rocksdb}, MongoDB~\cite{mongodb}, and CacheLib~\cite{cachelib}.
The first two are widely used log-structured storage backends, while CacheLib is
a high-performance caching engine. These case studies demonstrate the ease of
using \Shimmer and the performance benefits we get with each of the systems.
Here we compare \Shimmer with filesystem approaches like \fffs, and special-purpose
systems (\zenfs). Due to limited support for \zns \ssds, we could not include other 
filesystems in our comparison.

To evaluate \Shimmer, we set up a test server with 64-core AMD EPYC 7452 system
with \SI{128}{\giga\byte} of DRAM. We use Ubuntu 22.04 with Linux kernel 6.5,
and 2 \wesdig~\cite{wd_ultrastar_dc_zn540} each \SI{4}{\tera\byte} in size. We
used the latest stable release of each system and used unmodified bundled
benchmarking tools. Between runs of each benchmark, we issued zone reset and
\nvme format commands, and rebuilt the filesystems to ensure that each experiment
had a fresh start. Since \fffs required an in-place updatable region for
metadata, and \Shimmer uses it for \texttt{LOCK} files, we set up the region in
the \SI{4}{\giga\byte} random write space on the same drive.



\subsection{Case Study: RocksDB}

\begin{figure*}[htbp]
	\centering
	\includegraphics[width=\textwidth]{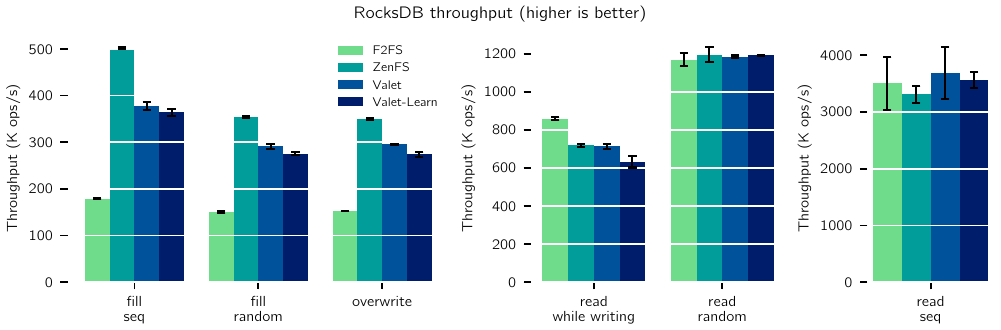}
	\caption{Throughput for \texttt{db\_bench} workloads.}
	\Description{Bar chart comparing throughput across six RocksDB db_bench workloads (fillseq, fillrandom, overwrite, readwhilewriting, readrandom, readseq) for four systems: F2FS, ZenFS, Valet, and Valet-Learn. Shows operations per second for each workload-system combination, with Valet and Valet-Learn generally achieving higher throughput than F2FS, and comparable performance to ZenFS.}
	\label{fig:rocksdb-throughput}
\end{figure*}

Evaluating RocksDB provides several benefits: it is widely used, it supports a
\zns-specific backend and can provide write stream hints. As \zenfs is
specifically tuned for RocksDB on \zns drives, it presents the gold standard for
the performance that highly tuned applications can achieve.  For our evaluation we ran
100 million operations, each with \SI{20}{\byte} keys and \SI{400}{\byte} values
for fill workloads (sequential and random), reads (sequential and random),
readwhilewriting, and overwrites. We used FIFO compaction as it provides natural
temporal separation to each of the systems.  For \fffs, we forwarded the write
hints provided by RocksDB.

We tested \Shimmer with two types of placement generation logic, the heuristic
approach and the automated approach. For the heuristic approach, we assigned
separate streams to each of the logs—the Write-Ahead Log (\textsc{wal}) and the
Sorted String Tables (\textsc{sst}). The placement groups were then based on the
timestamps that the extents filled up in. For automation, (\Shimmer-Learn) we
used batch-based mini KMeans clustering~\cite{10.1145/1772690.1772862}, a tuned
online approach that predicts affiliation between a specified number of
centroids. We then mapped the prediction of affinity to a particular centroid to
a stream.

\begin{figure*}[htbp]
	\centering
	\includegraphics[width=\textwidth]{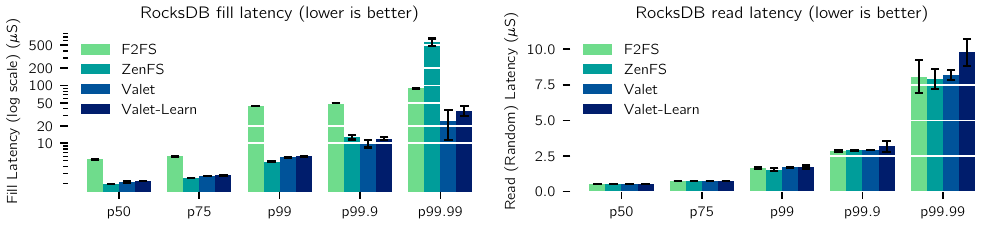}
	\caption{\Shimmer offers comparable latency to specifically tuned \zenfs and
		improved latency over \fffs.}
	\Description{Dual-panel line graph showing RocksDB latency percentiles (p50, p75, p99, p99.9, p99.99) for fillrandom and readrandom workloads. Compares F2FS, ZenFS, Valet, and Valet-Learn. F2FS and Valet achieve the lowest latencies across percentiles, while ZenFS shows significantly higher tail latency, especially at p99.9 and p99.99.}
	\label{fig:rocksdb-tail-latency}
\end{figure*}

As seen in~\cref{fig:rocksdb-throughput}, for inserts and updates, \Shimmer
offers over 2$\times$ improvement over \fffs and is comparable to tailored
approaches like \zenfs. While \zenfs offers higher throughput in each write
case, as we will see in~\cref{fig:rocksdb-tail-latency}, it suffers from high
tail latencies. For reads, \Shimmer equals any other approach: whether it be
\fffs using the kernel cache or \zenfs largely avoiding the kernel. In each case
\Shimmer-Learn suffers from a slight overhead as it generates hints on the fly.
However, unlike \zenfs and baseline \Shimmer, \Shimmer-Learn can be used with other
applications without any change. 

One of the main benefits of write isolation is the improvement in tail latency.
\Shimmer's user-space nature ensures that compared to the high cost of
kernel crossings and persistence, its impact is minimal. Here, \Shimmer is
comparable to \zenfs and offers a lower read latency than \fffs. \Shimmer-Learn
adds a small overhead, but it is still only about a dozen microseconds at 99.9
percentile, while still outperforming \fffs and \zenfs.

As seen in~\cref{fig:rocksdb-tail-latency}, \Shimmer offers better tail latency
over both \fffs and \zenfs for random inserts and random reads. This is
particularly evident at p99.99 (\Shimmer: $\sim$\SI{20}{\micro\second}, \zenfs:
$\sim$\SI{555}{\micro\second}), where \Shimmer benefits from the filesystem
optimizations in RocksDB, intercepting \texttt{readahead()} and pre-buffering
data, a filesystem optimization unavailable to \zenfs. \Shimmer sees a
~\SI{300}{\nano\second} impact on read latency due to differences between \fffs
and \zonefs, however since reads take several microseconds, it is unlikely to
have a real-world impact.  With automation, in \Shimmer-Learn we observe that it
can match the performance of hand-tuned approaches, however it can suffer small
tail latency spikes due to the
prediction engine.

\subsubsection{Why is \Shimmer faster?}

\begin{figure*}[htbp]
	\centering
	\includegraphics[width=\textwidth]{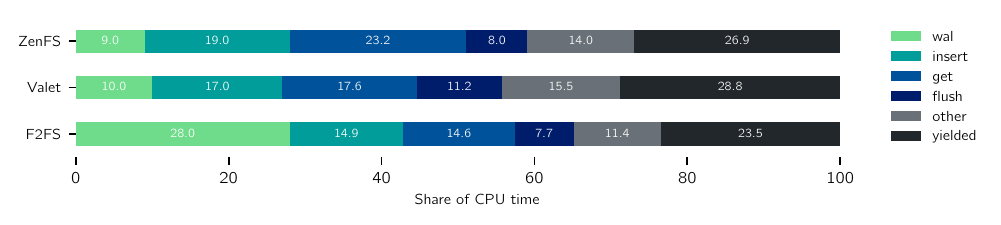}
	\caption[RocksDB benchmark breakdown]{Analyzing each system, we see how a
	filesystem like \fffs gets consumed by frequent flush operations.}
	\Description{Stacked bar chart showing CPU time breakdown for RocksDB
	operations across F2FS, ZenFS, and Valet systems. F2FS spends 28\% of CPU
	time on write-ahead log operations due to frequent syncs, while Valet and
	ZenFS spend only 9-10\% on WAL operations. Shows how filesystem overhead
	dominates F2FS performance.}
	\label{fig:rocksdb-breakdown}
\end{figure*}

\Shimmer provides intelligent hints discriminating between the frequent small
writes of the \textsc{wal} and the large writes of Sorted String Tables
(\textsc{sst}), utilizing separate device buffers to provide isolation.  Digging
deeper into the performance metrics (from \texttt{perf}~\cite{perf}) as seen in
\cref{fig:rocksdb-breakdown} for \fffs, the benchmark spends dozens of
milliseconds to persist the \textsc{wal}.  \textsc{wal} triggers locking in
\fffs, packing incoming streams across its six logs, performing metadata
updates, and allocating and freeing memory in the cache to support these
operations.

On the other hand, in \Shimmer the \textsc{wal} and the insert make up a
relatively small chunk of \cpu time, simply copying the buffer and periodically
persisting it. The userspace overhead is also limited, with these functions
accounting for less than 27\% of total program time, similar to \zenfs as
opposed to the 44\% of \fffs. Random reads (get operation) in \Shimmer are more
efficient than \zenfs, but add slight overhead over \fffs due to the extra steps
in logical block resolution, accounting for 17.5\% of the time as opposed to
23\% on \zenfs, which explains the latency spikes seen
in~\cref{fig:rocksdb-tail-latency}.


While RocksDB's composable nature makes it a great target for testing new
interfaces, one of the main benefits of \Shimmer's approach is generality, where
it works with more than just RocksDB, so we implemented placement for MongoDB.

\subsection{Case Study: MongoDB}

MongoDB's backend, WiredTiger~\cite{wiredtiger} gives users a choice for its
data structures: BTrees or Log-Structured Merge (\lsm) trees. As BTrees require
in-place updates, we focus on the \textsc{lsm} mode for \Shimmer. We tested
WiredTiger's performance with the medium-sized lsm tree test in \texttt{wt-perf}.
This included multithreaded reads, multithreaded updates, and simultaneous
multithreaded updates and reads.

WiredTiger performs two streams of writes: logs and \textsc{sst}s.  \Shimmer
does not need any changes to support the data, but we ran into issues supporting
the write-ahead logs.  WiredTiger uses \texttt{mmap()} writes for logging, which
necessitate in-place and out-of-order updates that are not supported in \vfs.
While \texttt{mmap()} databases are not recommended~\cite{crotty22-mmap💩}, this
limitation is handled by \Shimmer in its random-write region (with the lock
files). Fundamentally, mmap-writes do not guarantee ordering and are a violation
of the log-structured contract.  

For comparison, we looked at \fffs, providing it stream hints through \Shimmer
while \vfs uses \Shimmer's richer hints.

\begin{figure}[htbp]
	\centering
	\includegraphics[width=\columnwidth]{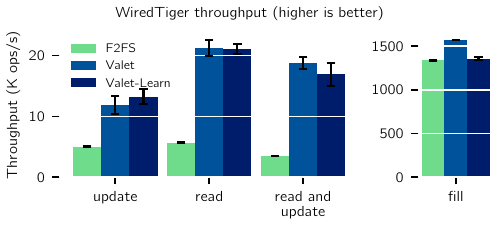}
	\caption{Due to physical separation of \texttt{log} and \texttt{.lsm} files
	to dedicated zones, \Shimmer can allow dramatically faster writes and
	updates in WiredTiger.}
	\Description{Bar chart comparing MongoDB WiredTiger throughput for four workloads (updateread, read, update, fill) across F2FS, Valet-Learn, and Valet systems. Shows operations per second, with Valet achieving significantly higher throughput than F2FS for write-heavy workloads (update and fill) due to physical separation of log and LSM files.}
	\label{fig:mongodb-throughput}
\end{figure}

In WiredTiger, we observe a dramatic improvement in updates and reads in the
\lsm mode. As we see in~\cref{fig:mongodb-throughput} \Shimmer with multithreaded
readers is more than 4$\times$ the throughput of \fffs, with write-heavy
workloads offering a 3$\times$ throughput. Again, the benefits come from the
simpler structure, offering dedicated write streams to different files
eliminates contention on the \ssd. \texttt{mmap}-writes when mixed with regular
writes can cause contention on filesystem resources in addition to the device
resources. \Shimmer still performs them, but assigning different device regions
reduces the contention. \Shimmer-Learn again presents a minimal overhead, making
a great case for exclusively using automated hints.

\begin{figure}[htbp]
	\centering
	\includegraphics[width=\columnwidth]{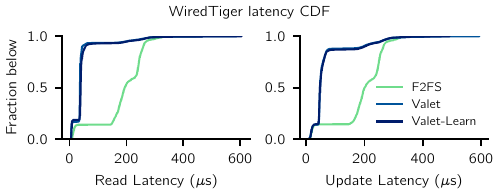}
	\caption{
		The isolation across logs eliminates contention and improves both read and
		update latency.}
	\Description{Dual-panel cumulative distribution function (CDF) graphs showing MongoDB latency for read and update operations. Compares F2FS, Valet-Learn, and Valet across latency values. Valet shows lower latency distributions for both operations, with steeper CDF curves indicating better tail latency due to isolation across logs eliminating contention.}
	\label{fig:mongodb-latency}
\end{figure}

In terms of latency, both instances of \Shimmer vastly outperform \fffs,
offering <\SI{100}{\micro\second} latency in almost all operations even as \fffs
exceeds \SI{200}{\micro\second}.

\subsection{Case Study: CacheLib}

\begin{figure}[htbp]
	\centering
	\includegraphics[width=\columnwidth]{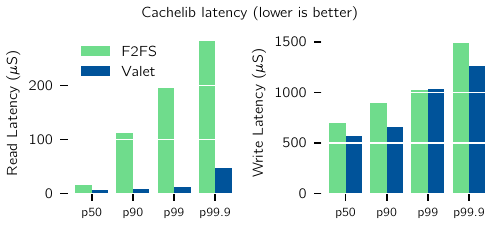}
	\caption{With CacheLib's \texttt{cachebench} tool, \Shimmer outperforms
		\fffs on throughput and read-write latency}
	\Description{Multi-panel graph showing CacheLib cachebench performance comparing F2FS and Valet. Shows throughput and read/write latency metrics. F2FS exhibits significantly higher read latency than Valet, while write latency remains comparable between the two systems.}
	\label{fig:cachelib}
\end{figure}

Finally, to demonstrate caching workloads we adopted Meta's caching library
CacheLib~\cite{258963} to \vfs, which required no change in \Shimmer's logic.
As we see in~\cref{fig:cachelib}, \Shimmer offers lower
latency for reads and writes.
To perform this test we set up CacheLib with \SI{128}{\mega\byte} of cache in
memory. We kept this cache minimal to accelerate spillover to flash, stress
testing our systems. We then performed 10 million get and set operations on the
cache.
CacheLib supports two kinds of caches: one optimized for small objects and one
for large. Again, we used heuristics to map these to different Streams. This is
one of the cases where automated approaches were identical, as the write streams
are identical.

On CacheLib \Shimmer sees a small (8\%) improvement in throughput \SI{56.8}
KOps/s as compared to \fffs' \SI{52.3} KOps/s but a large reduction in read
latency, particularly at p90 and above as seen in~\cref{fig:cachelib}.  This
small difference is partly due to the highly optimized Navy engine, which uses
optimizations \texttt{io\_uring}~\cite{axboe2019efficient}, skipping a lot of
filesystem overhead, and hence \Shimmer only sees a modest improvement. Reads
are still improved, as incoming writes do not block read operations.

\subsection{Multi-tenancy}

One of the main benefits of \zns \ssds is the ability to minimize degradation
caused by write interference from multiple streams of incoming writes. With
\Shimmer, multiple applications can be run in parallel on the same \ssd with a
significantly lower degradation in performance as their writes are isolated to
separate write resources. To evaluate degradation, we ran \Shimmer in a
multi-tenant environment, addressing MongoDB and RocksDB simultaneously on the
same device.  This results in 4 concurrent writers, each multithreaded: RocksDB,
RocksDB-\textsc{wal}, WiredTiger-\textsc{wal} and WiredTiger. We measured the
throughput and latency on both WiredTiger and RocksDB, focusing on tail latency,
which is typically adversely affected by such workloads. We observed that
throughput was not greatly affected on either system, but latency numbers show
interesting trends.

\begin{figure}[htbp]
    \centering
    \includegraphics[width=\columnwidth]{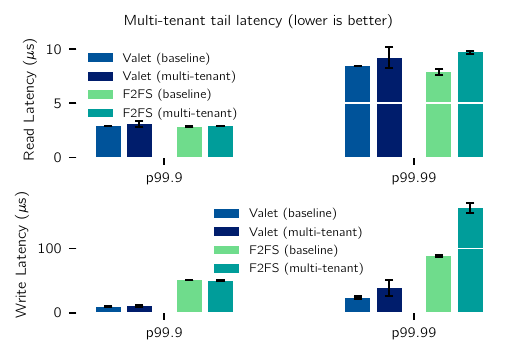}
    \caption{\Shimmer can exploit device-level isolation as \fffs gets bogged down by concurrent writers.}
    \Description{Four-panel graph showing read and write latency with and without interference for F2FS and Valet. Read latency remains similar across systems. Write tail latency for F2FS increases dramatically with interference, while Valet maintains low write tail latency due to device-level isolation across concurrent writers.}
    \label{fig:multi-latency}
\end{figure}

As we see in~\cref{fig:multi-latency}, tail latencies for writes see dramatic
spikes in filesystems like \fffs, owing to additional locking, and rigid
adherence to specific open zones based on the filesystem structure. This causes
the worst-case tail latency to spike almost 2$\times$. On the other hand,
\Shimmer can simply map writes to different device buffers and regions and they
see only a small degradation in latency. Despite being tailored for \zns \ssds,
systems like \zenfs inherently cannot allow multiple tenants or applications,
significantly limiting their data center applicability.

%


\subsection{Overhead}

Finally, to show that \Shimmer maintains these performance properties while
minimizing overhead, we measure the overhead in terms of data written and memory
usage.

\subsubsection{Write Amplification}

Reduced garbage collection and improved grouping help reduce write amplification
as data can be deleted together, avoiding the amplification caused by moving
data. As there is no on-device garbage collection, \vfs sees a device write
amplification factor of 1, similar to \fffs and \zenfs.
We ran an insert-heavy benchmark (100M sequential writes, random
writes, and updates). As we see in~\cref{table:garbage-collection}, in our
write-intensive experiments, \Shimmer frees up more space without moving any
data as \textsc{wal} blocks are freed up quickly, allowing \Shimmer to erase
without the need for relocating data. We excluded \zenfs from this comparison as
it does not implement garbage collection.

\begin{table}[htbp]
	\centering
	\caption{Garbage collection performance}
	\label{table:garbage-collection}
	\begin{tabular}{lrr}
		\toprule
		                         & \Shimmer              & \fffs                 \\\midrule
		Garbage Collection Calls & 8                     & 4                     \\
		Data Moved               & 0                     & \SI{1059}{\mebi\byte} \\
		Free Space at the End    & \SI{48.2}{\gibi\byte} & \SI{33.2}{\gibi\byte} \\\bottomrule
	\end{tabular}
\end{table}

As we see in \cref{table:garbage-collection}, freeing up space in \vfs was more
frequent and more efficient, moving no data as entire zones were invalidated due
to the better placement.

\subsubsection{Memory Usage}

\Shimmer in hint-only mode does not maintain any file data and uses no extra
memory. However, \vfs needs to perform write buffering, which requires
allocating memory depending on the number of open files. To speed up access,
\Shimmer maintains all its maps in memory, however except for \filemap, the rest are
either fixed size or frequently cleaned up. In our experiments we used
\SI{32}{\mega\byte} buffers with 2 write streams, totaling \SI{64}{\mega\byte}
write buffers which made up most of the actively used memory. This size is
smaller than the buffers maintained by filesystems in the kernel (\fffs memory
usage went up to \SI{200}{\mega\byte} in the same experiments).
Profiling the memory using KDE heaptrack~\cite{heaptrack}, \vfs used
\SI{73}{\mega\byte} at its peak in the evaluation experiments.


\subsubsection{Simplifying the storage stack}
\begin{table}[htbp]
	\centering
  \caption{Added lines of code}
  \label{tab:complexity}
  \begin{tabular}{l r r r}
		\toprule
  \textbf{System} & \textbf{Application} & \textbf{Kernel} & \textbf{Userspace} \\
  \midrule
  \zenfs & 4017 & 0 & 988 \\
  \fffs (\zns) & 0 & 38,188 & 1252 \\
  \Shimmer & 0 & 0 & 1700 \\\bottomrule
  \end{tabular}
  \end{table}

\Shimmer greatly simplifies the metadata and placement logic, replacing large
amounts of unnecessary kernel code with a simplified mapper that is a better fit
for modern applications and devices. As we discussed in~\cref{sec:why-not-fs},
added complexity in the kernel comes at a cost of security, and modifying
applications can make them hard to maintain. For instance, \zenfs already does
not work with the latest RocksDB.

\section{Related Work}\label{sec:related}


Early approaches to leveraging new \ssd interfaces involved modifying
applications~\cite{wang2014efficient,zhang2017flashkv} or using
application-specific backends.  \zenfs~\cite{bjorling2021zns, oh_zenfs_2023} is
a RocksDB plugin that maps RocksDB's files to \zns zones.
WALTZ~\cite{10.14778/3611479.3611495} further optimizes \zenfs, reducing tail
latency using the zone append command. 

\subsubsection*{Application-specific backends.}
Our work includes comparison of our approach with \zenfs, while it offers better
performance in certain conditions, \Shimmer is close behind and can allow
non-RocksDB applications. Additionally, \zenfs development has stalled,
highlighting challenges in maintaining application-specific backends.

Similarly, we attempted to compare \Shimmer's CacheLib performance with
Region-Cache~\cite{10.1145/3655038.3665946}, another \zns-based CacheLib
implementation that demonstrates the filesystem overhead for data-intensive
applications.  However, Region-Cache targets a CacheLib version from over a year
ago, with different kernel requirements, and dependencies, preventing direct
comparison in our test environment. In contrast, \Shimmer works with unmodified
CacheLib, avoiding version lock-in.  Overall application-specific tuning will
offer the best performance, but it requires a deep understanding of the
application as well as the storage medium, and significant engineering effort.
Even with these, it will suffer from version lock-in, and limited applicability.

\subsubsection*{Filesystems.}
Filesystems like \fffs~\cite{lee2015F2FS}, and \btrfs~\cite{rodeh2013btrfs}
offer \zns-specific improvements and are similar to \Shimmer with even broader
application compatibility: the ability to perform in-place updates.  \fffs is
perhaps the best example of a \zns-supporting filesystem, and we compare our
approach to theirs throughout this work.  Persimmon~\cite{10361006} is an
append-only fork of \fffs that requires no random-write area. Persimmon's
performance is similar to \fffs. However, it is tied to an older kernel version
(5.18), and hence, we were unable to evaluate it. \btrfs \zns support is
extremely limited and error-prone, and our benchmarks did not work on \btrfs.

Filesystem approaches face three key limitations that \Shimmer addresses:
(1)~in-kernel execution adds synchronization overhead, (2)~lack of
application-specific hint information limits data placement optimization, and
(3)~architectural constraints like F2FS's 3-data-log limit restrict parallelism.
\vfs, despite lacking filesystem features, can be used with applications that
rely on a filesystem interface while offering a much better write performance
and a comparable read performance.

\subsubsection*{Interposition.}
Interposition approaches outside the filesystem have used either
eBPF~\cite{zhong_xrp_nodate} or SPDK~\cite{yang2017spdk} as kernel-bypass
mechanisms. Other approaches have used
\sysint~\cite{noauthor_syscall_intercept_2023} to overload system calls for
persistent-memory programming by disassembling and patching binaries. The A few
\ldpreload-based filesystem prototypes exist, like PlasticFS~\cite{plasticfs}
and AVFS~\cite{avfs}, which allow peeking into compressed files.
Goanna~\cite{fast09valor} implemented a filesystem through ptrace extensions,
similar in spirit to \ldpreload. More recently, zIO~\cite{280938} used
user-space libraries using \ldpreload to eliminate unnecessary copies of data.
\Shimmer uses a similar interposition to redirect data.

\subsubsection*{Hint generation.}
While there are many proposals for hint formats, hint generation is largely an
overlooked area in \ssd research. Since workloads can be unpredictable, file
open provides few distinguishing features, and abstractions ensure that the
systems remain unaware of what they are storing, hint generation is an important
challenge~\cite{purandare2024enhancing}. We address it in \Shimmer with affinity
and lifetime, while traditional approaches such as the kernel hints have focused
on temperature.

\Shimmer uses interposition to inject hints and remap data if needed. No other
system interposes between the application and the filesystem in such a way.
Similar techniques have been implemented in different layers of the stack. Cloud
Storage Acceleration Layer~\cite{csal} enables the adoption of \zns with
clusters of varied storage and a host-based flash translation layer.  More
recently, Google demonstrated the use of filenames and open flags to decide data
placement, an approach similar to what \Shimmer-Learn
does~\cite{google_colossus_2025}. These approaches rely on diverse tiered
storage types with caches to balance random vs. sequential writes. \Shimmer
operates at the device interface level and could be the storage backend for such
services.

\Shimmer's modular architecture allows swapping hint generation strategies
through pluggable modules (Section~\ref{sec:hint-generation}) without affecting
other system components. This extensibility enables experimentation with
machine learning models, LLM-based classification, or domain-specific heuristics
as hint generation techniques evolve.


\section{Conclusion} \label{sec:conclusion}

The various efforts to introduce host-guided data placement present a sorry
picture: we see a recurring pattern of new interfaces being introduced,
demonstrated, and deprecated within a couple of years.  To change this paradigm,
we need to decouple the complexity of data placement from the applications and
filesystems, and make it easy to use these interfaces.  \Shimmer represents an
approach in this space, allowing filesystem and application development to be
unimpeded by changing interfaces, taking up the mantle to decide data placement
and device management.  It does this with rich hint interfaces and a composable
structure that allows flexibility across applications, operating systems, and
hardware protocols. In this paper, we demonstrated how application-specific
approaches were insufficient, and one-size-fits-all filesystems were inefficient.

Hardware is in turmoil. Compute, memory, and storage are dramatically changing
abstractions, making adoption  challenging while maintaining compatibility.
Ultimately, as is the case with \Shimmer, we believe that the way to address the
complexity is to isolate it in a dedicated layer that can be modified
independently of other parts.  This approach can speed up adoption of modern
interfaces, simplify programming, offer improved performance, and allow broad
application compatibility. \Shimmer can park your data more efficiently
for you.

\subsection*{Acknowledgements}
We are grateful to Matias Bjørling and Western Digital for the ZN540 drives and
feedback on this work. We would like to thank the industrial sponsors of the
Center for Research in Storage and Systems (CRSS), and the Kumar Malavalli
endowment at UC Santa Cruz for supporting this work. Feedback from Daniel
Bittman, Achilles Benetopoulos, Eugene Chou, and other CRSS researchers helped
shape \Shimmer, and we are grateful for their support. We would also like to
thank the anonymous reviewers for their feedback and suggestions, which made
this paper stronger.

\noindent The source code for the work presented in this paper is available at
\href{https://github.com/shimplify/valet}{https://github.com/shimplify/valet}.


\Urlmuskip=0mu plus 1mu\relax
\bibliographystyle{plain}
\bibliography{main}

\end{document}